\documentclass[twocolumn,prl]{revtex4}
\usepackage[T1]{fontenc}
\usepackage[latin1]{inputenc}
\usepackage{graphicx}

\begin{document}

\title[High Temperature Ca Vapor Cell]{A High Temperature Calcium Vapor Cell for Spectroscopy on the $4s^{2}$
$^{1}$S$_{0}$ to $4s4p$ $^{3}$P$_{1}$ Intercombination Line}

\author{Christopher J. Erickson, Brian Neyenhuis, Dallin S. Durfee}

\affiliation{Department of Physics and Astronomy}

\address{Brigham Young University, Provo, UT 84602}

\email{dallin_durfee@byu.edu}

\begin{abstract}
We have demonstrated a high temperature vapor cell for absorption
spectroscopy on the Ca intercombination line. The cell uses a dual
chamber design to achieve the high temperatures necessary for an optically
dense vapor while avoiding the necessity of high temperature vacuum
valves and glass-to-metal seals. We have observed over 50 percent
absorption in a single pass through the cell. Although pressure broadening
in the cell prevented us from performing saturated-absorption spectroscopy,
the broadening resulted in higher signal-to-noise ratios by allowing
us to probe the atoms with intensities much greater than the $0.2\mu\textrm{W}/\textrm{cm}^{2}$
saturation intensity of the unbroadened transition.
\end{abstract}
\maketitle

\section{Introduction}

Vapor cells are often a very convenient source of atoms for various
spectroscopic measurements. Although they suffer from relatively large
Doppler shifts compared to laser-cooled or beam sources, they are
typically very simple and inexpensive devices and can often produce
much larger optical densities than other more complicated schemes.
Vapor cells are widely used, and have been for some time. Most alkali-atom
based laser cooling experiments utilize vapor cells to lock the cooling
and probing lasers to the atomic transition. These vapor cells typically
consist of a glass vessel containing a small amount of alkali metal
which is heated with a low power heater or simply by contact with
the room-temperature environment to produce an optically thick vapor.

Alkaline-earth spectroscopy has gained considerable interest in the
past years, in part due to the use of the calcium $4s^{2}$ $^{1}$S$_{0}$
to $4s4p$ $^{3}$P$_{1}$ intercombination transition to realize
optical frequency standards. As such there has been interest in creating
simple sources of alkaline-earth vapors to use as references for these
experiments. In our case we wanted to have a relatively simply calcium
source for use in a laser lock for an excited state laser cooling
scheme and in order to find the frequencies of the modes of a high-stability
optical cavity relative to the calcium intercombination line.

Unfortunately the vapor pressures of the alkaline-earth elements are
considerably lower than those of the alkali atoms. Furthermore, the
narrowness of the intercombination line compared to the Doppler broadening
of a near-room temperature vapor results in a very small effective
absorption cross section, requiring high atomic densities in order
to achieve appreciable absorption. As such, producing an optically
dense vapor in a thermal cell requires vapors at temperatures which
are incompatible with standard vacuum technology.

There are many ways to achieve higher optical densities, such as the
use of an electric discharge \cite{Filho01ls,Hansen05df}. But these
schemes tend to be much more complicated than a simple vapor cell.
Even with these techniques it can still be difficult to produce the
needed densities for spectroscopy on the intercombination line. Before
deciding to construct a high-temperature calcium vapor cell we tried
to measure absorption on the intercombination line in a specially
designed high-density discharge cell but did not detect any significant
absorption. From measurements on the 423 nm resonance line we determined
that this cell produced a ground state density of $9\times10^{11}\textrm{atoms}/\textrm{cm}^{3}$.
In a single pass through the 10 cm discharge this should result in
less than 1\% absorption on the intercombination line.

\section{Vapor Cell Design}

When designing a vapor cell there are several key factors to consider.
The first is the length of the vapor cell. A longer cell will result
in higher absorption at a given temperature. We decided that we wanted
a relatively compact cell and decided to fix the length of the column
of vapor to be about 10 cm. As discussed below, the total length of
the vacuum chamber is a longer length of 51 cm to limit heat flow
away from the vapor and to allow the ends of the chamber to be cool
enough to mount to our optical table. To simplify the use of the cell
we decided to design the cell to achieve our target of 50\% absorption
in a single pass through the cell, although the windows in the cell
are large enough to accommodate multiple passes through the cell.

Given a cell length of $t=$10 cm and a target of 50\% absorption
we can calculate the required density of atoms in the cell using Beer's
law. Beer's law states that the fraction of light transmitted through
the cell when our laser is on resonance should be equal to $\exp(-n\sigma_{\textrm{eff}}t)$
where $n$ is the density of atoms and $t$ is the length of the cell.
The effective cross section, $\sigma_{\textrm{eff}}$, is lower than
the natural absorption cross section $\sigma_{0}=3\lambda^{2}/2\pi$
because, due to Doppler broadening, only a small fraction of the atoms
will be in resonance with the laser light at any given time. When
the laser is on resonance, only those atoms which have a Doppler shift
less than about one natural line width will scatter significant amounts
of light, and you would expect that the effective cross section should
be lower than the natural cross section by a factor of the order of
the ratio of the natural line width to the Doppler broadened line
width.

Because the Doppler broadening in our cell is much larger than both
the natural line width $\Gamma$ and the laser line width, the effective
on-resonance cross section $\sigma_{\textrm{eff}}$ is approximately
equal to $\sigma_{0}\Gamma\lambda(m/32\pi k_{B}T)^{1/2}$ where $\lambda$
is the resonant wavelength of the transition, $m$ is the mass of
an atom, $k_{B}$ is Boltzmann's constant, and $T$ is the temperature
of the vapor. In this limit $\sigma_{\textrm{eff}}$ does not depend
on the laser line width. Although the effective cross section is temperature
dependent, it is only mildly so --- it varies by only 18\% as the
vapor goes from $500^{\circ}\textrm{C}$ to $800^{\circ}\textrm{C}$.
As such, to estimate the required temperature we can calculate the
effective cross section at some reasonable temperature and then treat
it as a constant. For a calcium vapor at $500{}^{\circ}\textrm{C}$,
the effective on-resonance cross section for the intercombination
line is $8.8\times10^{-20}\textrm{m}^{2}$. With this cross section,
to get 10\% absorption in a column of vapor 10 cm long requires a
density of $1.2\times10^{13}$ atoms per cubic cm. This is accomplished
by a calcium vapor in equilibrium with solid calcium at $522^{\circ}\textrm{C}$.
For 50\% absorption, a density of $7.9\times10^{13}$ atoms per cubic
cm is needed, requiring a temperature of $587^{\circ}\textrm{C}$.

The construction of a vapor cell for spectroscopy on the calcium intercombination
transition is non-trivial due to the relatively high temperature required
and the scarcity of vacuum components which can operate at these temperatures.
Because the calcium will tend to accumulate at the coldest point in
the cell, the coldest point will eventually determine the vapor pressure
in the cell. In order to prevent fogging of the windows, the cell's
windows should be kept considerably hotter than the temperature of
the vapor. As such, producing a $587^{\circ}\textrm{C}$ vapor requires
portions of the cell to be considerably hotter than $587^{\circ}\textrm{C}$.
A blown glass cell would become soft and collapse at the required
temperatures. The most commonly used metal to metal or glass to metal
seals can only be baked up to about $450^{\circ}\textrm{C}$. The
only inexpensive and highly reliable way to make metal to metal vacuum
seals at these temperatures are to weld or braze components together.
In order to evacuate the chamber, a high temperature valve is also
required. Commercially available high-temperature valves are relatively
expensive and typically cannot be baked above $450^{\circ}\textrm{C}$.

To get around the expense and limitations of available high-temperature
vacuum components, we decided to use a two-chamber design. In this
design the hot calcium vapor is contained within an inner chamber
which is surrounded by an outer chamber. Because the vapor is contained
in the inner chamber, the entire outer chamber does not have to be
held at high temperatures. This allows simple commercially available
seals and valves to be used on the outer chamber. And because the
outer chamber is evacuated, the seals between the inner chamber and
outer chamber don't have to be vacuum tight in order to maintain vacuum
in the inner chamber. The only requirement on the seals between the
inner and outer chambers is that they leak calcium vapor at a sufficiently
low rate to allow the cell to operate for a long time without maintenance.

A scheme similar to our two-chamber design is discussed in \cite{Huang02ac}.
In this design there was no seal between the inner and outer chambers,
but long narrow tubes are used to limit calcium diffusion out of the
hot inner chamber. Because the inner chamber is evacuated through
these tubes no high temperature valve was required. Optical access
to the vapor was also done through these tubes, making window seals
unnecessary. The large diffusion rate of calcium out of the inner
chamber results in rapid fogging of the outer chamber windows. This
was overcome in \cite{Huang02ac} using mirrors inside the outer vacuum
to deflect the light to windows which were away from the direct line-of-sight
of the inner cell. These mirrors quickly coated with calcium, which
reflected the light reasonably well. We chose not to implement this
design for two reasons. First, we expect that calcium coated mirrors
would have a tendency to oxidize when the chamber is vented to reload
it with calcium. Second, and more critical, the diffusion rate of
calcium through these tubes is rather large --- in the tests reported
in \cite{Huang02ac} the cell ran for only 3 days running at 460$^{\circ}$C
using an unspecified amount of calcium. To avoid the trouble of continually
reloading the cell we designed a cell which uses windows to seal the
calcium into the inner chamber.

Our goal for the cell was to maintain a vapor dense enough to achieve
10\% absorption for over 1000 hours with an initial load of 5 grams
of calcium. This implies a maximum allowable leak rate which is equivalent
that of a hole with a cross section of about $1\,\textrm{mm}^{2}$.
While this is a very large hole by vacuum standards, it does require
some care to be taken. For example, the two 20-mm diameter windows
in our design have a combined circumference of 13 cm. If the average
distance between the edge of the windows and the mating surface is
greater than $8\,\mu\textrm{m}$, the calcium leak rate will be too
large. Looser requirements could be achieved simply by loading the
cell with more calcium. But this would increase the cost of operating
the cell and increase the risk that over time enough of the calcium
leaking out of the inner chamber would condense on the outer chamber
windows to significantly reduce their transparency.

\begin{figure}
\begin{center}\includegraphics[%
  width=8cm]{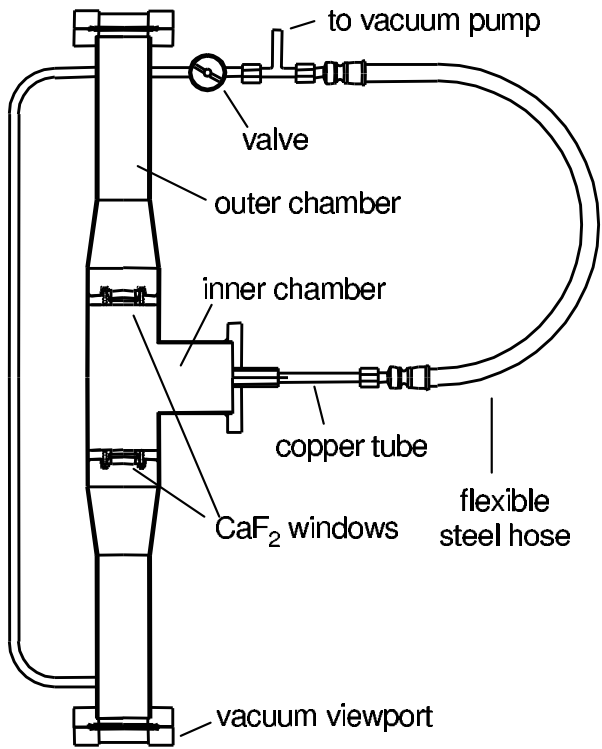}\end{center}

\caption{Diagram of the high temperature vapor cell. Diagram is to scale.\label{cap:diagram}}
\end{figure}

Our design is shown in Fig. \ref{cap:diagram}. In this design the
inner chamber is sandwiched between the two halves of the outer chamber.
The inner chamber is 10 cm in length, and 5.1 cm in diameter. This
diameter was chosen to accommodate the mounting of 20-mm diameter
windows on the ends of the inner chamber. Each half of the outer chamber
is 20 cm long to allow the far ends of the cell to be relatively cool
compared to the temperature of the inner chamber. The outer cell uses
standard off-the-shelf metal-to-metal flanges with copper gaskets,
and the windows on the outer cell are standard commercial vacuum windows.

To evacuate the cell, one half of the outer chamber is connected to
a vacuum pump through a valve. The two halves of the outer chamber
are connected by a 5/16 inch steel tube, allowing them to both be
evacuated at the same time. The inner chamber was connected to the
outer chamber through a long flexible steel tube connected to a short
piece of copper tubing which is brazed into a plate at the top of
the inner chamber (see Fig. \ref{cap:diagram}). After evacuating
the inner cell through this copper tube, we pinch the tube off using
a pinch-off tool commonly used by plumbers to rapidly seal water leaks
in copper tubing. We found that on a clean piece of copper this tool
creates a vacuum-tight seal as long as the tool remains on the tube.
To create a permanent seal, before removing the tool we cut the tube
off just above the pinch and fill the stub with silver solder. The
copper tube was made long enough to allow us to do the pinch-off procedure
several times before we have to braze a new tube into the cell.

Two windows separate the inner chamber from the two halves of the
outer chamber. To create a tight enough seal to contain the calcium
vapor, a copper gasket is placed between the windows and a flat steel
mating surface. The window is then compressed by a metal ring held
down with three bolts. The copper gaskets have beveled edges to create
a knife-edge contact on each side of the gasket, similar to the gaskets
used in the Kasevich seal \cite{Noble94uo} but with no attempt to
obtain a vacuum-tight junction. Sapphire windows were first used,
but even when held at a temperature over 100$^{\circ}$C hotter than
the calcium vapor they quickly fogged over (possibly due to a chemical
reaction). We eventually switched to calcium fluoride windows, which
have not shown any significant fogging and which have a thermal expansion
coefficient which is a much better match to stainless steel and copper.

The cell is heated by two band heaters at the location of the windows
and by a high temperature heat tape wrapped around a cylindrical aluminum
sleeve which is slid over the copper tube. The maximum temperatures
that the cell can maintain are set by the maximum rated temperatures
of the heaters. The band heaters can be heated as high as $800{}^{\circ}\textrm{C}$.
The heat tape is rated to go up to $740^{\circ}\textrm{C}$, sufficient
to produce atomic vapors with nearly 100\% absorption. If not wrapped
properly, however, heat tape can generate {}``hot spots'' and fail.
As such we ran the heat tape conservatively, well below the maximum
rated temperature.

The inner chamber is wrapped in several layers of high-temperature
fiberglass insulation and aluminum foil. The outer chamber is not
insulated, resulting in large temperature gradients such that the
far ends of the cell are at a temperature of less than $100^{\circ}$C
during operation of the cell. This allows us to seal the ends of the
cell with standard vacuum viewports and makes it possible to mount
the cell to our optics table without generating a significant heat
load. Since standard vacuum seals are used, we were able to easily
add a vacuum {}``T'' to one side of the outer chamber in order to
accommodate a vacuum gauge.

After loading calcium into the cell the cell is first heated using
just the two band heaters, ensuring that the windows are the hottest
parts of the inner chamber and thus preventing calcium build-up on
the windows. Eventually the calcium migrates to the colder copper
tube and the vapor density drops. After this migration occurs we typically
operate the cell with the band heaters fixed at $750^{\circ}$C and
use the heat tape to control the calcium vapor pressure.

\section{Results}

\begin{figure}
\begin{center}\includegraphics[%
  width=8cm,
  keepaspectratio]{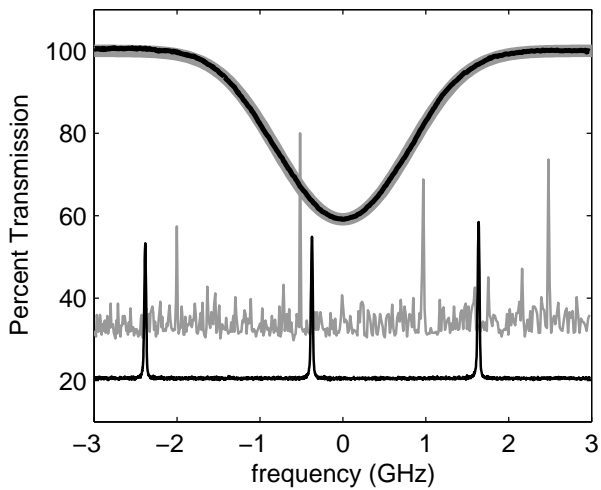}\end{center}

\caption{Absorption spectra. The black line at the top of the figure is the
measured transmission through the cavity as a function of the detuning
of the laser from the center of the intercombination line. It is drawn
over a thicker grey line which shows the calculated absorption for
a 10 cm long column of Ca vapor at a temperature of 750$^{\circ}$C
(the temperature of the hottest parts of the inner cell) but with
a density consistent with a solid source of calcium which is at 580$^{\circ}$C.
Superposed on this graph is the transmission through the reference
etalon (the black line with three sharp peaks at the bottom of the
graph) which we used to measure how far our laser had scanned. Above
the reference etalon trace in grey is the transmission through the
ultra-stable high-finesse cavity who's modes we measured relative
to the absorption line using our vapor cell. \label{cap:DataAndFit}}
\end{figure}

A typical absorption spectrum from our cell is shown in Fig. \ref{cap:DataAndFit}.
All of our absorption data was taken using a grating stabilized diode
laser of the design described in \cite{Merrill04it}. A portion of
the laser light was sent to a reference etalon to measure how far
the laser was scanned when taking absorption spectra. To verify the
calibration of the etalon's free-spectral range, we scanned the laser
frequency over a large range such that we could compare the etalon
to a commercial wavelength meter. This measurement agreed with the
manufacturer's specified free-spectral range of 2 GHz to within 0.5\%.

A beam splitter positioned just before the light entered the cell
directed a portion of the beam to a photo-diode so that we could record
a laser intensity reference to normalize away changes in the laser
intensity during the scan. Upon exiting the cell the light passed
through a pair of interference filters to block the blackbody radiation
from the cell. The light level was then detected on a photo-diode.
During each scan a digital oscilloscope recorded the light power before
the cell, the light power after the cell, and the amount of light
passing through the reference etalon. To remove stray-light background
levels, a second set of traces were recorded with the light blocked
near the laser.

To process the data we first subtracted the background from the scans
of the laser power before and after the cell. We then divided the
absorption signal at each point in time by the intensity reference.
Next we calculated a parabolic fit to the location in time of each
peak measured on the reference etalon and used this fit to convert
the time axis for each of our scans into a calibrated frequency axis
corrected for the nonlinear scan of the laser's frequency in time
(See Fig. \ref{cap:DataAndFit}). Finally, we corrected for etaloning
effects due to the windows in the cell by fitting the very edges of
the absorption signal, where essentially no absorption occurred, to
a second order polynomial. We subtracted the polynomial from the absorption
scan and then normalized the scan so that the leftmost edge was equal
to 1.

To test the longevity of the cell we loaded the cell with only 0.5
grams of calcium and ran the cell for over 60 hours at high enough
temperatures to produce between 30 and 50 percent absorption. Because
the rate at which calcium leaks out of the inner cell is proportional
to the density of the vapor, we can conclude that the amount of calcium
which leaked out of the inner chamber while we ran the cell is equivalent
to the leakage that would occur while operating the cell for over
200 hours at a lower temperature to produce 10\% peak absorption,
suggesting that we have surpassed our design goal of 1000 hours with
5 grams of calcium.

We have compared our absorption data to a model of the expected absorption
profile. We first started with a model which had just one free parameter,
the temperature of the vapor. Assuming that the temperature of the
solid calcium (which acts as the source for the calcium vapor) was
similar to the temperature of the vapor, we expected that this one
parameter would describe both the Doppler width of the absorption
curve as well as the density of the vapor (which, along with the Doppler
width, determines the on-resonance absorption). Although this simplest
model produced reasonably good results on our earliest data, after
running the cell for several hours the measured absorption curves
became noticeably wider than the curves predicted by our model.

The fact that our single-parameter model does not fit most of our
data well is not surprising. One would expect that once all of the
solid calcium had migrated into the copper tube, which is at a significantly
lower temperature than the rest of the inner chamber, a somewhat more
complicated two-parameter model would be needed. In this model the
temperature of the solid calcium, which acts as the source of the
vapor, and the effective temperature of the vapor were entered as
separate free parameters. We also tried fitting our data to a Voigt
profile model which included a third free parameter to include possible
effects of pressure broadening.

Our two-parameter model fits the data reasonably well when we assume
that the temperature of the vapor is close to the temperature of the
windows (see Fig. \ref{cap:DataAndFit}). Although we don't know the
precise temperature at the location of the solid calcium, from measurements
at the top and bottom of the copper tube we have a range of possible
temperatures. From the curve fits to our data it is apparent that
the density of the vapor is consistent with this range of temperatures.
When we add pressure broadening to our model it was very obvious that
the Voigt profile did not have the same shape as the data unless the
pressure broadening was assumed to be less than 100 MHz. This implies
that most of the disagreement between our simple one-parameter model
and the measured curves is due to the higher temperature of the vapor
rather than pressure broadening. Because 100 MHz is much smaller than
the Doppler width we can only ascertain an upper limit on pressure
broadening through curve fitting. We have determined a lower limit
on pressure broadening by observing an increase in the saturation
intensity; at the cell temperatures used to take the data in Fig.
\ref{cap:DataAndFit} we found that using all of the power from our
laser we could increase the intensity of our light to over 10,000
times the saturation intensity for a collision-free gas without seeing
any significant reduction in absorption, implying a lower limit on
the pressure broadening of several MHz.

Because the pressure-broadened width is significantly smaller the
Doppler width, a simpler absorption model which does not include pressure
broadening is still quite accurate. This makes data analysis much
faster, avoiding repetitive calculations of the integral in the Voigt
profile to optimise a curve fit. The simple model works because the
area under the absorption curve of a pressure broadened gas is independent
of the amount of broadening. This can be shown without any knowledge
of what the atoms are colliding with, what the collision frequency
is, or even whether the line shape is Lorentzian or if the impact
approximation applies. We only need to assume that collisions only
result in phase shifts with no direct excitation or de-excitation.
From the atom's frame it is equivalent to think of the laser field's
phase, rather than the atom's phase, as changing during the collision,
resulting in a light field which is effectively frequency broadened.
As such, pressure broadening can be thought of as the redistribution
of optical power to different frequencies (from the perspective of
the atom) with the total power in the light field remaining constant.

Pressure broadening in our cell is a disadvantage if precise location
of the atomic transition frequency is required. But for our purpose
of finding the frequencies of the modes of an ultra-stable high-finesse
cavity relative to the absorption line we only needed to find the
top of the absorption line to an accuracy of about 10 MHz. And because
the saturation intensity of collisionless calcium vapor is just 0.2$\mu$W/cm$^{2}$,
the increased saturation intensity due to collisions allows us to
send much more light through the cell and make huge improvements in
the signal-to-noise ratio. At these intensities our principle noise
source was the digitization noise of the digital oscilloscope which
we used to acquire data, and with little effort we acquired clean
enough data to find the center of the Doppler-broadened line within
the required 10 MHz. With our high-stability optical cavity we determined
that the center of the line did not shift measurably over the range
of temperatures used in the cell. For us the only disadvantage of
pressure broadening was the fact that we were unable to reach the
increased saturation intensity with our \textasciitilde{}1 mW laser
beam and were therefore unable to perform saturated-absorption spectroscopy.

\section{Conclusions}

We have demonstrated a high-temperature two-chambered vapor cell for
absorption spectroscopy on the calcium $4s^{2}$ $^{1}$S$_{0}$ to
$4s4p$ $^{3}$P$_{1}$ intercombination transition. We have achieved
high absorption and a long lifetime using a fairly simple and very
robust design, without needing to use expensive or complicated high
temperature seals or valves. Although the seals between the two chambers
do not need to be vacuum tight, we have found that care must be taken
in order to a keep the calcium loss rate reasonably small. Significant
pressure broadening in the cell has been used to our advantage, allowing
higher light intensities to be used, thereby improving the quality
of our data.

We would like to acknowledge the contributions of Scott Bergeson,
Wes Lifferth, and Rebecca Tang. This work was supported by a grant
from the Research Corporation.


\begin{thebibliography}{1}

\bibitem{Filho01ls}R. L. Cavasso-Filho, A. Mirage, A. Scalabrin, D. Pereira, and F. C.
Cruz, J. Opt. Soc. Am. B, \textbf{18}, 1922-1927 (2001).

\bibitem{Hansen05df}D. Hansen and A. Hemmerich, Phys. Rev. A, \textbf{72}, 022502
(2005).

\bibitem{Huang02ac}M.-S. Huanga, M.-H. Lu, and J.-T. Shy, Rev. Sci. Instrum., \textbf{73}, 3747-3749
(2002).

\bibitem{Noble94uo}A. Noble and M. Kasevich, Rev. Sci. Intrum., \textbf{65}, 3042-3-42 (1994).

\bibitem{Merrill04it}R. Merrill, R. Olson, S. Bergeson, and D.S. Durfee, Appl. Opt., \textbf{43}, 3910-3914 (2004).

\end{thebibliography}
\end{document}